\RequirePackage{fix-cm}
\documentclass[smallextended]{svjour3}       % onecolumn (second format)
\smartqed  % flush right qed marks, e.g. at end of proof
\usepackage[ruled]{algorithm2e} % For algorithms

\SetAlFnt{\small}
\SetAlCapFnt{\small}
\SetAlCapNameFnt{\small}
\SetAlCapHSkip{0pt}
\IncMargin{-\parindent}

\usepackage[justification=centering]{caption}
\usepackage{graphicx}
\usepackage{amsfonts}
\usepackage{amsmath}
\usepackage{multirow}
\usepackage{subfigure}
\usepackage{color}
\begin{document}

\title{Quantum Relief Algorithm%\thanks{Grants or other notes
}


\author{Wen-Jie Liu     \and
        Pei-Pei Gao     \and
        Wen-Bin Yu      \and
        Zhi-Guo Qu      \and
        Ching-Nung Yang    %etc.
}


\institute{
           W.-J. Liu
           \at Jiangsu Engineering Center of Network Monitoring, Nanjing University of Information Science \& Technology, Nanjing 210044, P.R.China
           \\
           \email{wenjiel@163.com}
           \and
           W.-J. Liu   \and
           P.-P. Gao   \and
           W.-B. Yu    \and
           Z.-G. Qu
           \at School of Computer and Software, Nanjing University of Information Science \& Technology, Nanjing 210044, P.R.China
            \and
           C.-N. Yang
           \at Department of Computer Science and Information Engineering, National Dong Hwa University, Hualien 974, Taiwan
 }

\date{Received: date / Accepted: date}

\maketitle

\begin{abstract}
Relief algorithm is a feature selection algorithm used in binary classification proposed by Kira and Rendell, and its computational complexity remarkable increases with both the scale of samples and the number of features. In order to reduce the complexity, a quantum feature selection algorithm based on Relief algorithm, also called quantum Relief algorithm, is proposed. In the algorithm, all features of each sample are superposed by a certain quantum state through the \emph{CMP} and \emph{rotation} operations, then the \emph{swap test} and measurement are applied on this state to get the similarity between two samples. After that, \emph{Near-hit} and \emph{Near-miss} are obtained by calculating the maximal similarity, and further applied to update the feature weight vector $WT$ to get $WT'$ that determine the relevant features with the threshold $\tau$. In order to verify our algorithm, a simulation experiment based on IBM Q with a simple example is performed. Efficiency analysis shows the computational complexity of our proposed algorithm is \emph{O(M)}, while the complexity of the original Relief algorithm is \emph{O(NM)}, where $N$ is the number of features for each sample, and $M$ is the size of the sample set. Obviously, our quantum Relief algorithm has superior acceleration than the classical one.
\end{abstract}

\section{Introduction}
\label{intro}
Machine learning refers to an area of computer science in which patterns are derived (``learned'') from data with the goal to make sense of previously unknown inputs. As part of both artificial intelligence and statistics, machine learning algorithms process large amounts of information for tasks that come naturally to the human brain, such as image recognition, pattern identification and strategy optimization.

Machine learning tasks are typically classified into two broad categories \cite{1}, supervised and unsupervised machine learning, depending on whether there is a learning ``signal" or ``feedback" available to a learning system. In supervised machine learning, the learner is provided a set of training examples with features presented in the form of high-dimensional vectors and with corresponding labels to mark its category. On the contrary, no labels are given to the learning algorithm, leaving it on its own to find structure in unsupervised machine learning. The core mathematical task for both supervised and unsupervised machine learning algorithm is to evaluate the distance and inner products between the high-dimensional vectors, which requires a time proportional to the size of the vectors on classical computers. As we know, it is ``curse of dimensionality'' \cite{2} to calculate the distance in high-dimensional condition. One of the possible solutions is the dimension reduction \cite{3}, and the other is the feature selection \cite{4}\cite{5}.

Relief algorithm is one of the most representative feature selection algorithms, which was firstly proposed by Kira et al. \cite{6}. The algorithm devises a ``relevant statistic'' to weigh the importance of the feature which has been widely applied in many fields, such as, hand gesture recognition \cite{7}, electricity price forecasting \cite{8} and power system transient stability assessment \cite{9}. Relief algorithm will occupy larger amount of computation resources while the number of samples and features become huger, which restricts the application of the algorithm.

Since the concept of quantum computer was proposed by the famous physicist Feynman \cite{10}, a number of remarkable outcomes have been proposed. For example, Deutsch's algorithms \cite{11}\cite{12} embody the superiority of quantum parallelism calculation, Shor's algorithm \cite{13} solves the problem of integer factorization in polynomial time, and Grover's algorithm \cite{14} has a quadratic speedup to the problem of conducting a search through some unstructured search space. With the properties of superposition and entanglement, quantum computation has the potential advantage for dealing with high dimension vectors which attract researchers to apply the quantum mechanics to solve some classical machine learning tasks such as quantum pattern matching \cite{15}, quantum probably approximately correct learning \cite{16}, feedback learning for quantum measurement \cite{17}, quantum binary classifiers \cite{18}\cite{19}, and quantum support vector machines \cite{20}.

Even in quantum machine learning, we still face with the trouble of ``curse of dimensionality'', thus dimension reduction or feature selection is a necessary preliminary before training high-dimensional samples. Inspired by the idea of computing the inner product between two vectors \cite{21}\cite{22}, we propose a Relief-based quantum parallel algorithm, also named quantum Relief algorithm, to effectively perform the feature selection.

The outline of this paper is as follows. The classic Relief algorithm is briefly reviewed in Sect. 2, the proposed quantum Relief algorithm is proposed in detail in Sect. 3, and a simulation experiment based on IBM Q with a simple example is given in Sect. 4. Subsequently, the efficiency of the algorithm is analyzed in Sect. 5, and the brief conclusion and the outlook of our work are discussed in the last section.

\section{Review of Relief algorithm}
\label{sec:2}
Relief algorithm is a feature selection algorithm used in binary classification (generalizable to polynomial classification by decomposition into a number of binary problems) proposed by Kira and Rendell \cite{6} in 1992. It is efficient in estimating features according to how well their values distinguish among samples that are near each other.

We can divide an $M$-sample set into two vector sets: $A{\rm{ = }}\left\{ {{v_j}{\rm{ | }}{v_j} \in {{\mathbb{R}}^{\rm{N}}}, j = 1,2, \cdots {M_1}} \right\}$, $B{\rm{ = }}\left\{ {{w_k}{\rm{ | }}{w_k} \in {{\mathbb{R}}^{\rm{N}}}, k = 1,2, \cdots {M_2}} \right\}$, where ${v_j}$, ${w_k}$ are $N$-feature samples: ${v_j} = {\left( {{v_{j1}},{v_{j2}}, \cdots ,{v_{jN}}} \right)^\mathsf{T}}$, ${w_k} = {\left( {{w_{k1}},{w_{k2}}, \cdots {w_{kN}}} \right)^\mathsf{T}}$, ${v_{j1}}, \cdots ,{v_{jN}},{w_{k1}}, \cdots ,{w_{kN}} \in \{ 0,1\} $, and the weight vector of $N$ features $WT = {\left( {w{t_1},w{t_2}, \cdots ,w{t_N}} \right)^\mathsf{T}}$ is initialized to all zeros. Suppose the upper limit of iteration is $T$, and the relevance threshold (that differentiate the relevant and irrelevant features) is $\tau (0\le \tau \le 1)$. The process of Relief algorithm is described in Algorithm~\ref{alg:one}.
% Algorithm Relief
\begin{algorithm}[t]
\SetAlgoNoLine
Init $WT = {\left( {0, \cdots ,0} \right)^\mathsf{T}}$. \\
\For{ t=1 to $T$}{
    Pick a sample $u$ randomly. \\
	Find the closest samples \emph{Near-hit} and \emph{Near-miss} in the two classes $A$ and $B$.\\
    \For {i=1 to $N$}{	
	   $WT\left[ i \right] = WT\left[ i \right]{\rm{ - }}\emph{diff}{\left( {i,u,\emph{Near-hit}} \right)^2}{\rm{ + }}\emph{diff}{\left( {i,u,\emph{Near-miss}} \right)^2}$.\\
    }
}
Select the most relevant features according to $WT$ and $\tau$.\\
\caption{Relief($A$, $B$, $T$, $\tau$)}
\label{alg:one}
\end{algorithm}

At each iteration, pick a random sample $u$, and then select the samples closest to $u$ (by $N$-dimensional Euclidean distance) from each class. The closest same-class sample is called \emph{Near-hit}, and the closest different-class sample is called \emph{Near-miss}. Update the weight vector such that,
%eq01
\begin{equation}
\label{eqn:01}
WT\left[ i \right] = WT\left[ i \right]{\rm{ - }}\emph{diff}{\left( {i,u,\emph{Near-hit}} \right)^2}{\rm{ + }}\emph{diff}{\left( {i,u,\emph{Near-miss}} \right)^2},
\end{equation}
where the function $\emph{diff}(i, u, v)$ is defined as below,
%eq02
\begin{equation}
\label{eqn:02}
diff(i,u,v) = \left\{
             \begin{array}{lcl}
             0{\kern 1pt} {\kern 1pt} {\kern 1pt} {\kern 1pt} {\kern 1pt} {\kern 1pt} {\kern 1pt} {\kern 1pt} {u_i} = {v_i} \\
             1{\kern 1pt} {\kern 1pt} {\kern 1pt} {\kern 1pt} {\kern 1pt} {\kern 1pt} {\kern 1pt} {\kern 1pt} {\kern 1pt} {\kern 1pt} {u_i} \ne {v_i}
             \end{array}
        .\right.
\end{equation}
Relief was also described as generalizable to polynomial classification by decomposition into a number of binary problems. However, as the scale of samples and the number of features increase, their efficiencies will drastically decline.

\section{Quantum Relief Algorithm}
\label{sec:3}
In order to handle the problem of large samples and large features, we propose a quantum Relief algorithm. Suppose the sample sets $A = \left\{ {{v_j},j = 1,2, \cdots {M_1}} \right\}$, $B = \left\{ {{w_k},k = 1,2, \cdots {M_2}} \right\}$, weight vector $WT$, the upper limit $T$ and threshold $\tau$ are the same as classical Relief algorithm defined in Sec. 2. Different from the classical one, all the features of each sample are represented as a quantum superposition state, and the similarity between two samples can be calculated in parallel. Algorithm~\ref{alg:two} shows the detailed procedure of quantum Relief algorithm.

% Algorithm QRelief
\begin{algorithm}[t]
\SetAlgoNoLine
Init $WT = {\left( {0, \cdots ,0} \right)^\mathsf{T}}$.\\
    Prepare the states for sample sets $A$ and $B$ through \emph{CMP} and \emph{rotation} operations,
    $\begin{array}{l}
{\left| {{\phi _A}} \right\rangle _j} = \frac{1}{{\sqrt N }}\left| j \right\rangle \sum\limits_{i = 0}^{N-1} {\left| {i} \right\rangle \left| 1 \right\rangle \left( {\sqrt {1 - {{\left| {{v_{ji}}} \right|}^2}} \left| 0 \right\rangle  + {v_{ji}}\left| 1 \right\rangle } \right)} \\
{\left| {{\phi _B}} \right\rangle _k} = \frac{1}{{\sqrt N }}\left| k \right\rangle \sum\limits_{i = 0}^{N-1} {\left| {i} \right\rangle \left| 1 \right\rangle \left( {\sqrt {1 - {{\left| {{w_{ki}}} \right|}^2}} \left| 0 \right\rangle  + {w_{ki}}\left| 1 \right\rangle } \right)}
\end{array}$.\\
\For{t=1 to $T$}{
    Select a state $\left| \phi  \right\rangle$ randomly from the state set $\left\{ {{{\left| {{\phi _A}} \right\rangle }_j}{\kern 1pt} } \right\}$ or ${\left\{ {{{\left| {{\phi _B}} \right\rangle }_k}{\kern 1pt} } \right\}}$.

    Perform a swap operation on $\left| \phi  \right\rangle$ to get $\left| \varphi  \right\rangle {\text{ = }}\frac{{\text{1}}}{{\sqrt N }}\left| l \right\rangle \sum\limits_{i = 0}^{N-1} {\left| i \right\rangle \left( {\sqrt {1 - {{\left| {{u_i}} \right|}^2}} \left| 0 \right\rangle  + {u_i}\left| 1 \right\rangle } \right)\left| 1 \right\rangle } $.\\

    Get $\begin{array}{l}
 {\left| {\left\langle {u} \mathrel{\left | {\vphantom {u {{v_j}}}} \right. \kern-\nulldelimiterspace} {{{v_j}}} \right\rangle } \right|^2} = \left( {1 - 2P_j^{l\left( A \right)}\left( 1 \right)} \right){N^2}\\
{\left| {\left\langle {u} \mathrel{\left | {\vphantom {u {{w_k}}}} \right. \kern-\nulldelimiterspace} {{{w_k}}} \right\rangle } \right|^2} = \left( {1 - 2P_k^{l\left( B \right)}\left( 1 \right)} \right){N^2}
\end{array}$ through \emph{swap test} and measurement operation.\\

    Obtain the maximum similarity: $\max \left\{ {{{\left| {\left\langle {u} \mathrel{\left | {\vphantom {u {{v_j}}}} \right. \kern-\nulldelimiterspace} {{{v_j}}} \right\rangle } \right|}^2}} \right\}$ and $\max \left\{ {{{\left| {\left\langle {u} \mathrel{\left | {\vphantom {u {{w_k}}}} \right. \kern-\nulldelimiterspace} {{{w_k}}} \right\rangle } \right|}^2}} \right\}$.\\

    \eIf{$u$ belongs to class $A$}{
        $\emph{Near-hit}={v_{\max }} , \emph{Near-miss}={\kern 1pt} {w_{\max }}$.\\
    }{
        $\emph{Near-hit}={w_{\max }}, \emph{Near-miss}={\kern 1pt} {v_{\max }}$.\\
    }

    \For{ i = 1 to $N$}{
    $w{t_i} = w{t_{i-1}} - \emph{diff}{\left( {i, u, \emph{Near-hit}} \right)^2} + \emph{diff}{\left({i, u, \emph{Near-miss}} \right)^2}$.\\
}
}

$\overline{WT}$ = $\left( {{1 \mathord{\left/ {\vphantom {1 T}} \right. \kern-\nulldelimiterspace} T}} \right)WT$.\\

\For {i = 1 to $N$}{
	\eIf{($\overline{WT}_i \ge \tau$)}{
        the \emph{i-th} feature is relevant.\\
    }{
        the \emph{i-th} feature is not relevant.\\
    }
    }
\caption{QRelief($A$, $B$, $T$, $\tau $)}
\label{alg:two}
\end{algorithm}

\subsection{State preparation}
\label{sec:3_1}
At the beginning of the algorithm, the classical information is converted into quantum superposition states. Quantum superposition state sets $\left\{ {{{\left| {{\phi _A}} \right\rangle }_j}{\kern 1pt} {\kern 1pt} {\kern 1pt} {\kern 1pt} |j = 1,2, \cdots ,{M_1}} \right\}$ and $\left\{ {{{\left| {{\phi _B}} \right\rangle }_k}{\kern 1pt} {\kern 1pt} {\kern 1pt} {\kern 1pt} |k = 1,2, \cdots ,{M_2}} \right\}$, which store all the feature values of ${v_j} \in A$ and ${w_k} \in B$, are prepared as below,
%Eq03
\begin{equation}
\label{eqn:03}
\begin{array}{l}
{\left| {{\phi _A}} \right\rangle _j} = \frac{1}{{\sqrt N }}\left| j \right\rangle \sum\limits_{i = 0}^{N-1} {\left| {i} \right\rangle \left| 1 \right\rangle \left( {\sqrt {1 - {{\left| {{v_{ji}}} \right|}^2}} \left| 0 \right\rangle  + {v_{ji}}\left| 1 \right\rangle } \right)} \\
{\left| {{\phi _B}} \right\rangle _k} = \frac{1}{{\sqrt N }}\left| k \right\rangle \sum\limits_{i = 0}^{N-1} {\left| {i} \right\rangle \left| 1 \right\rangle \left( {\sqrt {1 - {{\left| {{w_{ki}}} \right|}^2}} \left| 0 \right\rangle  + {w_{ki}}\left| 1 \right\rangle } \right)}
\end{array},
\end{equation}
where ${v_{ji{\kern 1pt}}}$ and ${w_{ki{\kern 1pt}}}$ represent the \emph{i-th} feature value of vector ${v_{j{\kern 1pt} }}$ and ${w_{k{\kern 1pt} }}$, respectively. Suppose the initial state is $\left| j \right\rangle {\left| 0 \right\rangle ^{ \otimes n}}\left| 1 \right\rangle \left| 0 \right\rangle $ ($n = \left\lceil {{{\log }_2}\left( {N } \right)} \right\rceil $) and we want to prepare the state ${\left| {{\phi _A}} \right\rangle _j}{\kern 1pt} {\kern 1pt} $, its construction process consists of the following three steps.

\begin{figure}
  \centering
  \includegraphics[width=2.7in]{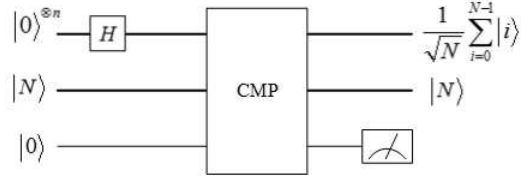}
  \caption{The circuit of quantum operation performed on ${\left| 0 \right\rangle ^{ \otimes n}}$ to obtain the state ${{\rm{1}} \over {\sqrt N }}\sum\limits_{i = 0}^{N - 1} {\left| i \right\rangle }$ when the measurement outcome is 0. Here, the thick horizontal line represents multiple qubits, while the thin horizontal line represents one qubit, and \emph{H} is the Hadamard gate, and CMP can be implemented using the circuit in Fig. 2.}
  \label{fig:1}
\end{figure}

First of all, the \emph{H} (i.e., Hadamard gate) and \emph{CMP} operations are performed on ${\left| 0 \right\rangle ^{ \otimes n}}$ to obtain the state ${\raise0.7ex\hbox{$1$} \!\mathord{\left/
 {\vphantom {1 {\sqrt N }}}\right.\kern-\nulldelimiterspace}
\!\lower0.7ex\hbox{${\sqrt N }$}}\sum\limits_{i = 0}^{N-1} {\left| i \right\rangle } $,
%eq04
\begin{equation}
\label{eqn:04}
{\left| 0 \right\rangle ^{ \otimes n}} \xrightarrow{\emph{H} {\kern 2pt} and {\kern 2pt}\emph{CMP}{\kern 2pt} operations}{{\rm{1}} \over {\sqrt N }}\sum\limits_{i = 0}^{N-1} {\left| i \right\rangle }.
\end{equation}
Fig.~\ref{fig:1} depicts the detailed circuit of these operations, where the \emph{CMP} operation is a key component which is used to tailor the state ${\raise0.7ex\hbox{$1$} \!\mathord{\left/ {\vphantom {1 {\sqrt {{2^n}} }}}\right.\kern-\nulldelimiterspace} \!\lower0.7ex\hbox{${\sqrt {{2^n}} }$}}\sum\limits_{i = 0}^{{2^n-1}} {\left| i \right\rangle } $ into the target state ${\raise0.7ex\hbox{$1$} \!\mathord{\left/ {\vphantom {1 {\sqrt N }}}\right.\kern-\nulldelimiterspace}\!\lower0.7ex\hbox{${\sqrt N }$}}\sum\limits_{i = 0}^{N-1} {\left| i \right\rangle } $, and its definition is
%eq05
\begin{equation}
\label{eqn:05}
CMP\left| i \right\rangle \left| N \right\rangle \left| 0 \right\rangle  = \left\{ {
\begin{array}{*{20}{c}}
{\left| i \right\rangle \left| N \right\rangle \left| 0 \right\rangle ,i < N}\\
{\left| i \right\rangle \left| N \right\rangle \left| 1 \right\rangle ,i  \ge  N}
\end{array}} .\right.
\end{equation}
After the \emph{CMP} operation, the quantum state which greater than $N$ (i.e., the last qubit is $\left| 1 \right\rangle $) will be clipped off.
%figure1 quantum CMP operation
%figure2  CMP operation
\begin{figure}
  \centering
  \includegraphics[width=2.2in]{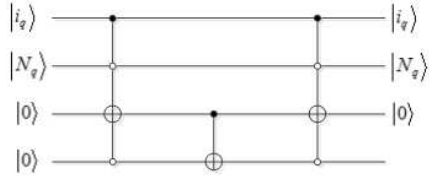}
  \caption{The circuit for performing CMP illustrated for a single qubit inputs $\left| {{i_q}} \right\rangle $ and $\left| {{N_q}} \right\rangle $. Here, the third line is the auxiliary qubit, while the last one is the result qubit which will contain $\left| {1} \right\rangle $ if $i \ge N$, and be $\left| {0} \right\rangle $ if $i < N$.}.
  \label{fig:2}
\end{figure}

And then, an unitary rotation operation
%eq06
\begin{equation}
\label{eqn:06}
{R_y}\left( {2{{\sin }^{ - 1}}{v_{ji}}} \right) = \left[ {
\begin{array}{*{20}{c}}
  {\sqrt {1 - {{\left| {{v_{ji}}} \right|}^2}} }&{ - {v_{ji}}} \\
  {{v_{ji}}}&{\sqrt {1 - {{\left| {{v_{ji}}} \right|}^2}} }
\end{array}} \right]
\end{equation}
is performed on the last qubit to obtain ${\left| {{\phi _A}} \right\rangle _j}{\kern 1pt} {\kern 1pt}$,
%eq07
\begin{equation}
\label{eqn:07}
{{\rm{1}} \over {\sqrt N }}\left| j \right\rangle \sum\limits_{i = 0}^{N-1} {\left| {i} \right\rangle \left| 1 \right\rangle \left| 0 \right\rangle } \buildrel {{R_y}} \over
 \longrightarrow {1 \over {\sqrt N }}\left| j \right\rangle \sum\limits_{i = 0}^{N-1} {\left| {i} \right\rangle \left| 1 \right\rangle \left( {\sqrt {1 - {{\left| {{v_{ji}}} \right|}^2}} \left| 0 \right\rangle  + {v_{ji}}\left| 1 \right\rangle } \right)} .
\end{equation}

\subsection{Similarity calculation}
\label{sec:3_2}
The similarity is a coefficient that describes how close two samples are, and it is obviously inversely related to the Euclidean distance. After the state preparation, the main work of this phase is to calculate the similarity between $\left| \phi  \right\rangle$ and other states in state sets $\left\{ {{{\left| {{\phi _A}} \right\rangle }_j}{\kern 1pt} } \right\}$ and ${\left\{ {{{\left| {{\phi _B}} \right\rangle }_k}{\kern 1pt} } \right\}}$, where $\left| \phi  \right\rangle$ is a state selected randomly from $\left\{ {{{\left| {{\phi _A}} \right\rangle }_j}{\kern 1pt} } \right\}$ or ${\left\{ {{{\left| {{\phi _B}} \right\rangle }_k}{\kern 1pt} } \right\}}$. For simplicity, suppose $\left| \phi  \right\rangle$ is the \emph{l-th} state from $\left\{ {{{\left| {{\phi _A}} \right\rangle }_j}{\kern 1pt} } \right\}$,
%eq08
\begin{equation}
\label{eqn:08}
\left| \phi  \right\rangle {\text{ = }}\frac{1}{{\sqrt N }}\left| l \right\rangle \sum\limits_{i = 0}^{N-1} {\left| {i} \right\rangle \left| 1 \right\rangle \left( {\sqrt {1 - {{\left| {{u_i}} \right|}^2}} \left| 0 \right\rangle  + {u_i}\left| 1 \right\rangle } \right)} ,
\end{equation}
which corresponds to the chosen sample $u$ from sample set $A$ in the classical scenario. The detailed process is as follows.

First, a swap operation is performed on the last two qubits of $\left| \phi  \right\rangle$ to obtain a new state,
%eq09
\begin{equation}
\label{eqn:09}
\left| \varphi  \right\rangle  = \frac{1}{{\sqrt N }}\left| l \right\rangle \sum\limits_{i = 0}^{N-1} {\left| {i} \right\rangle \left( {\sqrt {1 - {{\left| {{u_i}} \right|}^2}} \left| 0 \right\rangle  + {u_i}\left| 1 \right\rangle } \right)\left| 1 \right\rangle } .
\end{equation}

\begin{figure}
\centering
\includegraphics[width=2.0in]{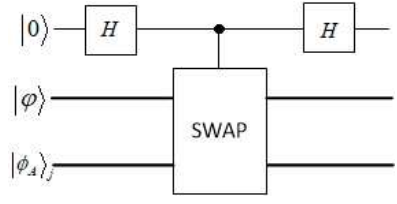}
\caption{The circuit of \emph{swap test} operation. Here \emph{H} is the Hadamard gate, and $\bullet$ represents the control qubit conditional being set to one.}
\label{fig:3}
\end{figure}

Second, the \emph{swap test} \cite{23} operation (its circuit is given in Fig.~\ref{fig:3}) is applied on $\left( {\left| \varphi  \right\rangle ,{\kern 1pt} {\kern 1pt} {\kern 1pt} {\kern 1pt} {\kern 1pt} {\kern 1pt} {{\left| {{\phi _A}} \right\rangle }_j}} \right)$,
%eq10
\begin{equation}
\label{eqn:10}
\begin{array}{l}
\left| 0 \right\rangle \left| \varphi  \right\rangle {\left| {{\phi _A}} \right\rangle _j}\xrightarrow{swap{\kern 1pt} {\kern 1pt} {\kern 1pt} {\kern 1pt} test}\left[ {{1 \over 2}\left| 0 \right\rangle (\left| \varphi  \right\rangle {{\left| \phi_A  \right\rangle }_j} + {{\left| \phi_A  \right\rangle }_j}\left| \varphi  \right\rangle ) + {1 \over 2}\left| 1 \right\rangle (\left| \varphi  \right\rangle {{\left| \phi_A  \right\rangle }_j} - {{\left| \phi_A  \right\rangle }_j}\left| \varphi  \right\rangle )} \right]\\
\end{array}.
\end{equation}
%Figure 3:Swap test
If we measure the first qubit with $\left| 1 \right\rangle \left\langle 1 \right| \otimes I \otimes I$, the probability of measurement result to be $\left| {\rm{1}} \right\rangle $ is
%eq11
\begin{equation}
\label{eqn:11}
\begin{array}{l}
{P_j^l}(1) = \left[ {{1 \over 2}\left\langle 0 \right|(\left\langle \varphi  \right|{{\left\langle \phi_A  \right|}_j} + {{\left\langle \phi_A  \right|}_j}\left\langle \varphi  \right|) + {1 \over 2}\left\langle 1 \right|(\left\langle \varphi  \right|{{\left\langle \phi_A  \right|}_j} - {{\left\langle \phi_A  \right|}_j}\left\langle \varphi  \right|)} \right]\left| 1 \right\rangle \left\langle 1 \right| \otimes I \otimes I\\

\qquad \qquad\left[ {{1 \over 2}\left| 0 \right\rangle (\left| \varphi  \right\rangle {{\left| \phi_A  \right\rangle }_j} + {{\left| \phi_A  \right\rangle }_j}\left| \varphi  \right\rangle ) + {1 \over 2}\left| 1 \right\rangle (\left| \varphi  \right\rangle {{\left| \phi_A  \right\rangle }_j} - {{\left| \phi_A  \right\rangle }_j}\left| \varphi  \right\rangle )} \right]\\

\;\,\qquad= \left[ {{1 \over 2}\left\langle 1 \right|(\left\langle \varphi  \right|{{\left\langle \phi_A  \right|}_j} - {{\left\langle \phi_A  \right|}_j}\left\langle \varphi  \right|)} \right]\left| 1 \right\rangle \left\langle 1 \right| \otimes I \otimes I\left[ {{1 \over 2}\left| 1 \right\rangle (\left| \varphi  \right\rangle {{\left| \phi_A  \right\rangle }_j} - {{\left| \phi_A  \right\rangle }_j}\left| \varphi  \right\rangle )} \right]\\

\;\,\qquad= {1 \over 2} - {1 \over 2}{\left| {{{\left\langle {\varphi }
 \mathrel{\left | {\vphantom {\varphi  \phi }}
 \right. \kern-\nulldelimiterspace}
 {\phi_A } \right\rangle }_j}} \right|^2}
\end{array},
\end{equation}
As we know, the inner product between $\left| \varphi  \right\rangle $ and prepared state ${\left| {{\phi _A}} \right\rangle _j}$ can be calculated as below,
%eq12
\begin{equation}
\label{eqn:12}
\begin{array}{l}
{\left\langle {\varphi }
 \mathrel{\left | {\vphantom {\varphi  {{\phi _A}}}}
 \right. \kern-\nulldelimiterspace}
 {{{\phi _A}}} \right\rangle _j} = {1 \over N}\sum\limits_i {(u_i)^*{v_{ji}}}  = {1 \over N}\left\langle {u}
 \mathrel{\left | {\vphantom {u {{v_j}}}}
 \right. \kern-\nulldelimiterspace}
 {{{v_j}}} \right\rangle\\
\end{array}.
\end{equation}

From Eqs. (\ref{eqn:11}) and(\ref{eqn:12}), we can get the similarity between samples $u$ and $v_j$,
%eq13
\begin{equation}
\label{eqn:13}
\begin{array}{l}
{\left| {\left\langle {u} \mathrel{\left | {\vphantom {u {{v_j}}}} \right. \kern-\nulldelimiterspace}
 {{{v_j}}} \right\rangle } \right|^2} = \left( {1 - 2P_j^{l\left( A \right)}\left( 1 \right)} \right){N^2}\\
\end{array}.
\end{equation}

Finally, we can find out the max-similarity sample ${v_{\max }} \in A$ through the classical maximum traversal search algorithm among the set $ \left\{ {{{\left| {\left\langle {u} \mathrel{\left | {\vphantom {u {{v_j}}}} \right. \kern-\nulldelimiterspace} {{{v_j}}} \right\rangle } \right|}^2}} \right\}$.

Through the above method, we can also find out the other max-similarity sample ${w_{\max }} \in B$. When having finished all these steps, the algorithm turns to the next phase.

\subsection{Weight vector update}
\label{sec:3_3}
The first step is to determine the closest same-class sample \emph{Near-hit} and the different-class sample \emph{Near-miss}, which obey the following rule,
%eq14
\begin{equation}
 \label{eqn:14}
\left\{ {\begin{array}{*{20}{c}}
  {\emph{Near-hit} = {v_{\max }},{\kern 1pt} {\kern 1pt}  \emph{Near-miss} = {w_{\max }}}&{ {\kern 1pt} {\kern 1pt} if{\kern 1pt} {\kern 1pt} u \in A} \\
  {\emph{Near-hit} = {w_{\max }},{\kern 1pt} {\kern 1pt} \emph{Near-miss} = {v_{\max }}}&{{\kern 1pt}  {\kern 1pt} if{\kern 1pt} {\kern 1pt} u \in B}
\end{array}} .\right.
 \end{equation}
After determining \emph{Near-hit} and \emph{Near-miss}, we can update every element of weight vector $WT = {\left( {w{t_1},w{t_2}, \cdots ,w{t_N}} \right)^\mathsf{T}}$ with them,
 %eq15
\begin{equation}
\label{eqn:15}
w{t_i} = w{t_{i - 1}} - \emph{diff}{\left( {i, u, \emph{Near-hit}} \right)^2} + \emph{diff}{\left( {i, u, \emph{Near-miss}} \right)^2}{\kern 1pt} {\kern 1pt},
\end{equation}
where $1\le i \le N$.

\subsection{Feature selection}
\label{sec:3_4}
After iterating the similarity calculation and weight vector update $T$ runs, the algorithm jumps out of the loop with the final vector $WT$ as output result. The remainder of the algorithm is to select the ``real" relevant features.

We firstly divide $WT$ by $T$, and obtain its mean vector $\overline{WT}$,
%eq16
\begin{equation}
\label{eqn:16}
\overline{WT} = \frac{1}{T}WT.
\end{equation}
Then, we select relevant features according to the preset threshold $\tau$. To be specific, those features are selected if their corresponding values in $\overline{WT}$ are greater than $\tau$ and discarded in the opposite case,
%eq17
\begin{equation}
\label{eqn:17}
\left\{ {\begin{array}{*{20}{c}}
{the{\kern 2pt} \emph{i-th}{\kern 2pt} feature{\kern 2pt} is{\kern 2pt} relevant {\kern 36pt} if{\kern 3pt}\overline{WT}_i \ge \tau }\\
{the{\kern 2pt} \emph{i-th}{\kern 2pt} feature{\kern 2pt} is{\kern 2pt}NOT{\kern 2pt} relevant {\kern 8pt} if{\kern 3pt}\overline{WT}_i < \tau }
\end{array}} .\right.
\end{equation}

\section{Example and its experiment}
\label{sec:4}
\begin{figure}
\centering
\includegraphics[width=2.0in]{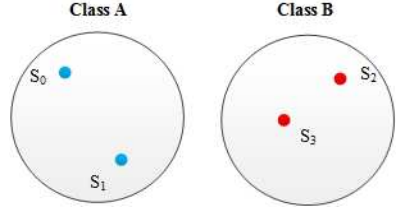}
\caption{The simple example with four samples in classes $A$ and $B$.}
\label{fig:4}
\end{figure}

Suppose there are four samples(see Tab.~\ref{tab:1}), ${S_0} = (1,0,1,0)$, ${S_1} = (1,0,0,0)$, ${S_2} = (0,1,1,0)$, ${S_3} = (0,1,0,0)$, thus the $n$ in Eq. (\ref{eqn:04}) is 2, and they belong to two classes: $A = \{ {S_0},{S_1}\}$, $B = \{ {S_2},{S_3}\}$, which is illustrated in Fig.~\ref{fig:4}.
%table 1
\begin{table}
\centering
\caption{The feature values of four samples. Each row represents all the feature values of a certain sample, while each column denotes a certain feature value of all the samples.}
\label{tab:1}
\begin{tabular}{cllll}
\hline\noalign{\smallskip}
 & $F_0$ & $F_1$ & $F_2$ & $F_3$  \\
\noalign{\smallskip}\hline\noalign{\smallskip}
$S_0$ & 1 & 0 & 1 & 0\\
$S_1$ & 1 & 0 & 0 & 0\\
$S_2$ & 0 & 1 & 1 & 0\\
$S_3$ & 0 & 1 & 0 & 0\\
\noalign{\smallskip}\hline
\end{tabular}
\end{table}
%Figure 4:sample image

First, the four initial quantum states are prepared as follows,
%eq18
\begin{equation}
\label{eqn:18}
\left\{ {\begin{array}{*{20}{c}}
{\left| \psi  \right\rangle _{{{\rm{S}}_{\rm{0}}}}} = \left| {{\rm{00}}} \right\rangle {\left| {\rm{0}} \right\rangle ^{ \otimes 2}}\left| 1 \right\rangle \left| 0 \right\rangle \\
{\left| \psi  \right\rangle _{{{\rm{S}}_{\rm{1}}}}} = \left| {{\rm{01}}} \right\rangle {\left| {\rm{0}} \right\rangle ^{ \otimes 2}}\left| 1 \right\rangle \left| 0 \right\rangle \\
{\left| \psi  \right\rangle _{{{\rm{S}}_{\rm{2}}}}} = \left| {{\rm{10}}} \right\rangle {\left| {\rm{0}} \right\rangle ^{ \otimes 2}}\left| 1 \right\rangle \left| 0 \right\rangle \\
{\left| \psi  \right\rangle _{{{\rm{S}}_{\rm{3}}}}} = \left| {{\rm{11}}} \right\rangle {\left| {\rm{0}} \right\rangle ^{ \otimes 2}}\left| 1 \right\rangle \left| 0 \right\rangle
\end{array}} .\right.
\end{equation}
Taking ${\left| \psi  \right\rangle _{{{\rm{S}}_{\rm{0}}}}}$ as an example, the ${H^{ \otimes 2}}$ operation is applied on the third and fourth qubits,
%eq19
\begin{equation}
\label{eqn:19}
\begin{array}{l}
\left| {{\rm{00}}} \right\rangle {\left| {\rm{0}} \right\rangle ^{ \otimes 2}}\left| 1 \right\rangle \left| 0 \right\rangle \xrightarrow{H^{ \otimes 2}}{1 \over 2}\left| {{\rm{00}}} \right\rangle \sum\limits_{i = 0}^3 {\left| i \right\rangle \left| 1 \right\rangle \left| 0 \right\rangle }
\end{array}.
\end{equation}
Then we perform ${R_y}$ rotation (see Eq. (\ref{eqn:06})) on the last qubit, where
%eq20
\begin{equation}
\label{eqn:20}
\left\{ {\begin{array}{*{20}{c}}
{R_y}(2{\sin ^{ - 1}}{v_{00}}) = {R_y}(2{\sin ^{ - 1}}{v_{02}})  =  -iY = \left[ {
\begin{array}{*{20}{c}}
   0 & { - 1}  \cr
   1 & 0  \cr
\end{array} } \right]\\
{R_y}(2{\sin ^{ - 1}}{v_{01}}) = {R_y}(2{\sin ^{ - 1}}{v_{03}}) $$ = I= \left[ {
\begin{array}{*{20}{c}}
   1 & 0  \cr
   0 & 1  \cr
\end{array} } \right]
\end{array}} ,\right.
\end{equation}
and can get
%eq21
\begin{equation}
\label{eqn:21}
{\left| \phi \right\rangle _{{{\rm{S}}_{\rm{0}}}}}={1 \over 2}\left| {{\rm{00}}} \right\rangle \sum\limits_{i = 0}^3 {\left| i \right\rangle \left| 1 \right\rangle \left( {\sqrt {{\rm{1 - }}{{\left| {{v_{0i}}} \right|}^2}} \left| 0 \right\rangle  + {v_{0i}}\left| 1 \right\rangle } \right)} .
\end{equation}
The other quantum states are prepared in the same way and they are listed as below,
%eq22
\begin{equation}
\label{eqn:22}
\left\{ {\begin{array}{*{20}{c}}
{\left| \phi  \right\rangle _{{S_0}}} = {1 \over 2}\left| {00} \right\rangle \sum\limits_{i = 0}^3 {\left| i \right\rangle } \left| 1 \right\rangle \left( {\sqrt {1 - {{\left| {{v_{0i}}} \right|}^2}} \left| 0 \right\rangle  + {v_{0i}}\left| 1 \right\rangle } \right)\\
{\left| \phi  \right\rangle _{{S_1}}} = {1 \over 2}\left| {01} \right\rangle \sum\limits_{i = 0}^3 {\left| i \right\rangle } \left| 1 \right\rangle \left( {\sqrt {1 - {{\left| {{v_{1i}}} \right|}^2}} \left| 0 \right\rangle  + {v_{1i}}\left| 1 \right\rangle } \right)\\
{\left| \phi  \right\rangle _{{S_2}}} = {1 \over 2}\left| {10} \right\rangle \sum\limits_{i = 0}^3 {\left| i \right\rangle } \left| 1 \right\rangle \left( {\sqrt {1 - {{\left| {{v_{2i}}} \right|}^2}} \left| 0 \right\rangle  + {v_{2i}}\left| 1 \right\rangle } \right)\\
{\left| \phi  \right\rangle _{{S_3}}} = {1 \over 2}\left| {11} \right\rangle \sum\limits_{i = 0}^3 {\left| i \right\rangle } \left| 1 \right\rangle \left( {\sqrt {1 - {{\left| {{v_{3i}}} \right|}^2}} \left| 0 \right\rangle  + {v_{3i}}\left| 1 \right\rangle } \right)
\end{array}} .\right.
\end{equation}

Second, we randomly select a sample (assume ${\left| \phi  \right\rangle _{{{\rm{S}}_{\rm{0}}}}}$ is that one), and perform similarity calculation with other samples (i.e., ${\left| \phi  \right\rangle _{{{\rm{S}}_{\rm{1}}}}}$, ${\left| \phi  \right\rangle _{{{\rm{S}}_{\rm{2}}}}}$, ${\left| \phi  \right\rangle _{{{\rm{S}}_{\rm{3}}}}}$). Taking ${\left| \phi  \right\rangle _{{{\rm{S}}_{\rm{0}}}}}$ and ${\left| \phi  \right\rangle _{{{\rm{S}}_{\rm{1}}}}}$ as an example, we perform a \emph{swap} operation between the last two qubits of ${\left| \phi  \right\rangle _{{{\rm{S}}_{\rm{0}}}}}$,
%eq23
\begin{equation}
\label{eqn:23}
\begin{array}{l}
{\left| \phi  \right\rangle _{{{\rm{S}}_{\rm{0}}}}}\xrightarrow{swap}\left| {{\varphi}} \right\rangle = {1 \over 2}\left| {{\rm{00}}} \right\rangle \sum\limits_{i = 0}^3 {\left| i \right\rangle  \left( {\sqrt {{\rm{1 - }}{{\left| {{v_{0i}}} \right|}^2}} \left| 0 \right\rangle  + {v_{0i}}\left| 1 \right\rangle } \right)\left| 1 \right\rangle}
\end{array}.
\end{equation}
After that, the \emph{swap test} operation is applied on ($\left| {{\varphi}} \right\rangle $, $\left| {{\phi}} \right\rangle_{{{\rm{S}}_{\rm{1}}}} $),
%eq24
\begin{equation}
\label{eqn:24}
\begin{array}{l}
\left| 0 \right\rangle \left| \varphi  \right\rangle {\left| {{\phi}} \right\rangle_{{{\rm{S}}_{\rm{1}}}}}\xrightarrow{swap{\kern 1pt} {\kern 1pt} {\kern 1pt} {\kern 1pt} test}{1 \over 2}\left| 0 \right\rangle \left( {\left| \varphi  \right\rangle {{\left| {{\phi}} \right\rangle_{{{\rm{S}}_{\rm{1}}}} }} +  {{\left| {{\phi}} \right\rangle_{{{\rm{S}}_{\rm{1}}}} }}\left| \varphi  \right\rangle} \right) + {1 \over 2}\left| 1 \right\rangle \left( {{\left| \varphi  \right\rangle{\left| {{\phi}} \right\rangle_{{{\rm{S}}_{\rm{1}}}} }}  - {{\left| {{\phi}} \right\rangle_{{{\rm{S}}_{\rm{1}}}} }}\left| \varphi  \right\rangle } \right)\\
\end{array},
\end{equation}
then, we measure the result shown in Eq.(\ref{eqn:24}) and obtain the probability of the first qubit being $\left| {\rm{1}} \right\rangle $ is
%eq25
\begin{equation}
\label{eqn:25}
\begin{array}{l}
{P_1^0}(1) = {1 \over 2} - {1 \over 2}{\left| {{{\left\langle {\varphi }
 \mathrel{\left | {\vphantom {\varphi  \phi }}
 \right. \kern-\nulldelimiterspace}
 {\phi } \right\rangle_{{{\rm{S}}_{\rm{1}}}} }}} \right|^2}
\end{array}.
\end{equation}
In terms of Eq. (\ref{eqn:12}), the inner product between $\left| {{\varphi}} \right\rangle $ and $\left| {{\phi}} \right\rangle_{S_1}$ can be represented as
%eq26
\begin{equation}
\label{eqn:26}
\left\langle {\varphi }
 \mathrel{\left | {\vphantom {\varphi  \phi }}
 \right. \kern-\nulldelimiterspace}
 {\phi } \right\rangle_{S_1}  = {1 \over 4}\sum\limits_i {{S_0}_i^*{S_{1i}} = {1 \over 4}\left\langle {{{S_0}}}
 \mathrel{\left | {\vphantom {{{S_0}} {{S_1}}}}
 \right. \kern-\nulldelimiterspace}
 {{{S_1}}} \right\rangle }.
\end{equation}
From Eqs. (\ref{eqn:25}) and (\ref{eqn:26}), the similarity between $\left| {{S_0}} \right\rangle$ and $\left| {{S_1}} \right\rangle$ is
%eq27
\begin{equation}
\label{eqn:27}
{\left| {\left\langle {{{S_0}}}
 \mathrel{\left | {\vphantom {{{S_0}} {{S_1}}}}
 \right. \kern-\nulldelimiterspace}
 {{{S_1}}} \right\rangle } \right|^2} = 16(1 - 2{P_1^0}(1)),
\end{equation}
here, the value of ${P_1^0}(1)$ can be determined by measurement.

\begin{figure}[htbp]
\centering
\includegraphics[width=4.5in]{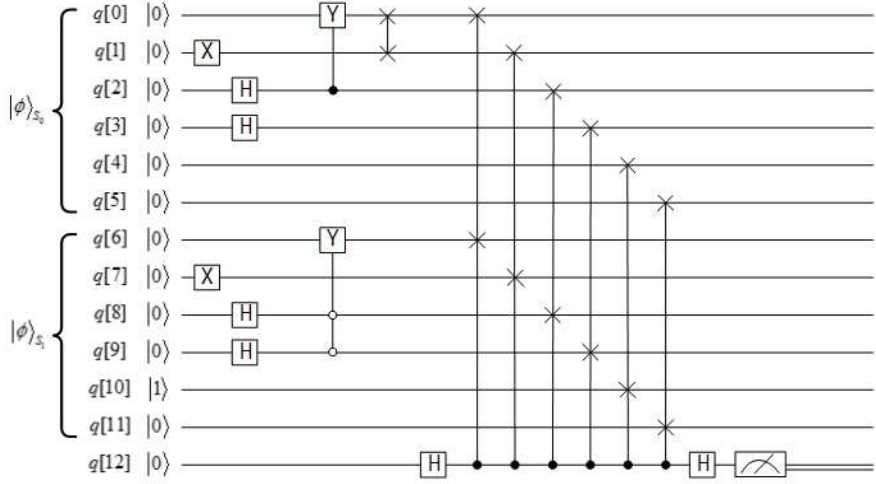}
\caption{One of the ideal experiment circuits of QRelief algorithm running on IBM Q platform. $q[0]-q[5]$ represents the randomly selected quantum state $\left| {{\phi}} \right\rangle_{S_0}$, $q[6]-q[11]$ represents $\left| {{\phi}} \right\rangle_{S_1} $, and $q[12]$ is the result qubit. \emph{H} is the Hadamard gate, \emph{X}, \emph{Y} are Pauli-X, Pauli-Y gates, the symbol of two crosses connected by a line represents the \emph{swap} operation, $\circ $ represents the control qubit conditional being set to zero, and $\bullet$ represents the control qubit conditional being set to one.}
\label{fig:5}
\end{figure}

In order to obtain the measurement result and also verify our algorithm, we choose the IBM Q [24] to perform the quantum processing (Fig. \ref{fig:5} gives the schematic diagram of our experimental circuit)\footnote{In the experiment, we program our algorithm based on the QISKit toolkit[25] and phyton language, and \textcolor[rgb]{0.00,0.00,1.00}{remotely connect the online IBM QX5 device} to execute quantum processing.}. After the experiment, we can get ${P_1^0}(1)$ which is shown in Tab.~\ref{tab:2}.
%Figure 5:Experiment circuit of QRelief based on IBM Q platform
%table 2: P(1)
\begin{table}
\centering
\caption{The probabilities $P_j^l(1)$ of the first qubit being $\left| {\rm{1}} \right\rangle $}
\label{tab:2}       % Give a unique label
\begin{tabular}{clll}
\hline\noalign{\smallskip}
 Iteration times($T$)& $u$ & $Sample$ & $P_j^l(1)$   \\
\noalign{\smallskip}\hline\noalign{\smallskip}
\multirow{3}*{1} & \multirow{3}*{$S_0$} & $S_1$ & 0.49023438  \\
&  & $S_2$ & 0.49902344  \\
&  & $S_3$ & 0.49121094  \\
\multirow{3}*{2} & \multirow{3}*{$S_1$} & $S_0$ & 0.50097656  \\
&  & $S_2$ & 0.52246094  \\
&  & $S_3$ & 0.53417969  \\
\multirow{3}*{3} & \multirow{3}*{$S_2$} & $S_0$ & 0.50683594  \\
&  & $S_1$ & 0.50878906  \\
&  & $S_3$ & 0.49218750  \\
\multirow{3}*{4} & \multirow{3}*{$S_3$} & $S_0$ & 0.49804688  \\
&  & $S_1$ & 0.49218750  \\
&  & $S_2$ & 0.50195312  \\
\noalign{\smallskip}\hline
\end{tabular}
\end{table}

According to ${P_1^0}(1)$ (from Tab. \ref{tab:2}) and Eq. (\ref{eqn:27}), we can calculate the similarity between $\left| {{S_0}} \right\rangle$ and ${\left| S_1  \right\rangle}$,
%eq28
\begin{equation}
\label{eqn:28}
\begin{array}{l}
{\left| {\left\langle {{{S_0}}} \mathrel{\left | {\vphantom {{{S_0}} {{S_1}}}} \right. \kern-\nulldelimiterspace} {{{S_1}}} \right\rangle } \right|^2} = 16(1 - 2{P_1^0}(1))\\
\;\;\;\qquad\qquad=16(1-2*0.49023438)\\
\;\;\;\qquad\qquad \approx 0.3125
\end{array}.
\end{equation}
In the same way, the other two similarities ($\left| {{S_0}} \right\rangle$, ${\left| S_2  \right\rangle}$),  ($\left| {{S_0}} \right\rangle$, ${\left| S_3  \right\rangle}$) can also be obtained (which are illustrated in Tab. \ref{tab:3}).
%table3:Similarities between samples
\begin{table}
\centering
\caption{Similarities between samples}
\label{tab:3}       % Give a unique label
\begin{tabular}{cllll}
\hline\noalign{\smallskip}
 & $S_0$ & $S_1$ & $S_2$ & $S_3$  \\
\noalign{\smallskip}\hline\noalign{\smallskip}
$S_0$ & $-$ & 0.3125 & 0.03125 & 0.28125 \\
$S_1$&   & $-$ & 0.71875 & 1.09375  \\
$S_2$&   &   & $-$ & 0.25  \\
$S_3$&  &   &   & $-$  \\
\noalign{\smallskip}\hline
\end{tabular}
\end{table}

\begin{figure}
\centering
\includegraphics[width=2.0in]{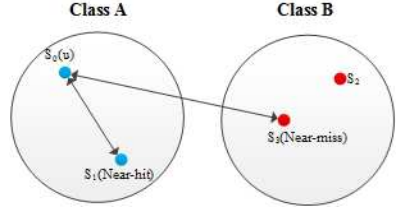}
\caption{Finding \emph{Near-hit} and \emph{Near-miss}.}
\label{fig:6}
\end{figure}

Third, From Tab. \ref{tab:3}, it is easy to find \emph{Near-hit} is $S_1$ and \emph{Near-miss} is $S_3$ (as shown in Fig. \ref{fig:6}). Then, the weight vector is updated by applying Eq.(\ref{eqn:15}), and the result of $WT$ is listed in the second row of Tab. \ref{tab:4} for the first iteration.

%Figure 6:fig_Nearest samples
%table 3: The update result of $WT$
\begin{table}
\centering
\caption{The update result of $WT$}
\label{tab:4}
\begin{tabular}{cc}
\hline\noalign{\smallskip}
$Iteration times(T)$ & $Weight vector(WT)$  \\
\noalign{\smallskip}\hline\noalign{\smallskip}
1 & [1  1  0 0]    \\
2 & [2  2 -1 0]   \\
3 & [3  3 -1 0]   \\
4 & [4  4 -2 0]   \\
\noalign{\smallskip}\hline
\end{tabular}
\end{table}

The algorithm iterates $T$ times (in our example, $T$=4) as above steps, and obtains all the $WT$ results shown in Tab. \ref{tab:4}. After $T$-th iterations, $WT=[4,4,-2,0]$, then $\overline{WT}=[1,1,-1/2,0]$. Since the threshold $\tau=0.5 $, so the selected features are $F_0$ and $F_1$, i.e., the first and second features.
\section{Efficiency analysis}
\label{sec:5}
In classical Relief algorithm, it takes \emph{O(N)} times to calculate the Euclidean distance between $u$ and each of the $M$ samples, the complexity of finding its \emph{Near-hit} and \emph{Near-miss} is related to $M$, and the loop iterates $T$ times, so the computational complexity of classical Relief algorithm is \emph{O(TMN)}. Since $T$ is a constant which affects the accuracy of relevance levels, but it is chosen independently of $M$ and $N$, so the complexity can be simplified to \emph{O(MN)}. In addition, an $N$-dimensional vector in the Hilbert space is represented by $N$ bits in the classical computer, and there are $M$ samples (i.e., $M$ $N$-dimensional vectors) needed to be stored in the algorithm, so the classical Relief algorithm will consume $O(MN)$ bits storage resources.

In our quantum Relief algorithm, all the features of each sample are superposed on a quantum state ${\left| {{\phi _A}} \right\rangle _j}$ or ${\left| {{\phi _B}} \right\rangle _k}$, then the similarity calculation between two states, which is shown in Eq. (\ref{eqn:13}), is just required to be taken $O(1)$ time. As same as Relief algorithm, the similarity between the selected state $\left| \varphi  \right\rangle$ and each state in $\left\{ {\left| {{\phi _A}} \right\rangle } \right\}$, $\left\{ {\left| {{\phi _B}} \right\rangle } \right\}$ is calculated, taking $O(M)$ times, to obtain \emph{Near-miss} and \emph{Near-hit}, and the loop iterates $T$ times, so the proposed algorithm totally takes \emph{O(TM)} times. Since $T$ is a constant, the computational complexity can be rewritten as \emph{O(M)}. On the viewpoint of resource consumption, each quantum state in state sets $\left\{ {\left| {{\phi _A}} \right\rangle _j} \right\}$ and $\left\{ {\left| {{\phi _B}} \right\rangle _k} \right\}$ is represented as $${\left| {{\phi _A}} \right\rangle _j} = \frac{1}{{\sqrt N }}\left| j \right\rangle \sum\limits_{i = 0}^{N-1} {\left| {i} \right\rangle \left| 1 \right\rangle \left( {\sqrt {1 - {{\left| {{v_{ji}}} \right|}^2}} \left| 0 \right\rangle  + {v_{ji}}\left| 1 \right\rangle } \right)}$$ or $${\left| {{\phi _B}} \right\rangle _k} = \frac{1}{{\sqrt N }}\left| k \right\rangle \sum\limits_{i = 0}^{N-1} {\left| {i} \right\rangle \left| 1 \right\rangle \left( {\sqrt {1 - {{\left| {{w_{ki}}} \right|}^2}} \left| 0 \right\rangle  + {w_{ki}}\left| 1 \right\rangle } \right)},$$ and it consists of $O({\log _2}N)$ qubits. Since $j = 1,2, \cdots ,{M_1}$, $k = 1,2, \cdots ,{M_2}$ and $M={M_1}+{M_2}$, that means that there are such $M$ quantum states needed to be stored in our algorithm, so it will consume $O(MlogN)$ qubits storage resources.

Tab.~\ref{tab:5} illustrates the efficiency comparison, including the computational complexity and resource consumption, between classical Relief and our quantum Relief algorithms. As shown in the table, the computational complexity of our algorithm is $O(M)$, which is obviously superior than $O(MN)$ in classical Relief algorithm. On the other hand, the resource that our algorithm needs to consume is $O(MlogN)$ qubits, while the classical Relief algorithm consumes $O(MN)$ bits.
% table 4
\begin{table}
\centering
\caption{Efficiency comparison between classical Relief and quantum Relief algorithms }
\label{tab:5}
\begin{tabular}{lll}
\hline\noalign{\smallskip}
 & Complexity & resource consumption  \\
\noalign{\smallskip}\hline\noalign{\smallskip}
Relief Algorithm &$O(MN)$ &$O(MN)$ bits\\
 Quantum Relief Algorithm& $O(M)$ &$O(MlogN)$ qubits\\
\noalign{\smallskip}\hline
\end{tabular}
\end{table}

\section{Conclusion and Discussion}
\label{sec:6}
With quantum computing technologies nearing the era of commercialization and quantum supremacy, recently machine learning seems to be considered as one of the "killer" applications. In this study, we utilize quantum computing technologies to solve a simple feature selection problem (just used in binary classification), and propose the quantum Relief algorithm, which consist of four phases: state preparation, similarity calculation, weight vector update, and features selection. Furthermore, we verify our algorithm by performing quantum experiments on the IBM Q platform. Compared with the classical Relief algorithm, our algorithm holds lower computation complexity and less resource consumption (in terms of number of resource). To be specific, the complexity is reduced from $O(MN)$ to $O(M)$, and the consumed resource is shortened from $O(MN)$ bits to $O(MlogN)$ qubits.

Although this work just focuses on the feature selection problem (even the simple binary classification), but the method can be generalized to implement the other Relief-like algorithms, such as ReliefF \cite{26}, RReliefF \cite{27}. Besides, we are interested in utilizing quantum technologies to deal with some classic high-dimension massive data processing, such as text classification, video stream processing, data mining, computer version, information retrieval, and bioinformatics.

\begin{acknowledgements}
The authors would like to thank the anonymous reviewers and editor for their comments that improved the quality of this paper. This work is supported by the National Nature Science Foundation of China (Grant Nos. 61502101, 61501247 and and 61672290), the Natural Science Foundation of Jiangsu Province, China (Grant No. BK20171458), the Six Talent Peaks Project of Jiangsu Province, China (Grant No. 2015-XXRJ-013), the Natural Science Foundation for Colleges and Universities of Jiangsu Province, China (Grant No. 16KJB520030), the Practice Innovation Training Program Projects for Jiangsu College Students (Grant No. 201810300016Z), and the Priority Academic Program Development of Jiangsu Higher Education Institutions (PAPD).
\end{acknowledgements}


\end{document}